\documentclass[preprint]{elsarticle}

\usepackage{amssymb}
\usepackage{subfigure}
\usepackage{layout}
\usepackage[T1]{fontenc}
\usepackage[cp1250]{inputenc}

\usepackage{amsmath}
\usepackage{graphicx}
\usepackage{amssymb}

\usepackage{pstricks}
\usepackage{latexsym}

\begin{document}

\begin{frontmatter}

\title{Information channel capacity in the field theory estimation}

\author[rvt]{J.~S{\l}adkowski \corref{cor1} 
}
\author[focal]{J.~Syska \corref{cor2} 
}

\cortext[cor1]
{Corresponding author: J.~S{\l}adkowski ({\it Email address:} jan.sladkowski@us.edu.pl)}
\cortext[cor2]
{Corresponding author: J.~Syska ({\it Email address:} jacek.syska@us.edu.pl)}


\address[rvt]{Department of Astrophysics and Cosmology, Institute of Physics, University of 
Silesia, Uniwersytecka 4, 40-007 Katowice, Poland}
\address[focal]{Department of Field Theory and Particle Physics, 
Institute of Physics, University of 
Silesia, Uniwersytecka 4, 40-007 Katowice, Poland}

\begin{abstract}
The construction of the information capacity for the vector position parameter in the Minkowskian space-time is presented. This lays the statistical foundations of the kinematical term of the Lagrangian of the physical action for many field theory models,  derived by the extremal physical information method of Frieden and Soffer.
\end{abstract}

\begin{keyword}
channel information capacity \sep Fisher information \sep Stam's information \sep causality 
\end{keyword}

\end{frontmatter}

\section{Introduction}

\label{Introduction}

The Fisher information ($I_{F}$) is a second degree covariant tensor,  
which is one of the contrast functions defined on the statistical space ${\cal S}$ \cite{Amari Nagaoka book}. 
It is
the local version of the Kulback-Leibler entropy for two distributions 
infinitesimally displaced from one another 
\cite{Frieden,Bengtsson_Zyczkowski}. 
This, in turn, is the model selection tool used in the standard maximum likelihood (ML) method 
and the basic notion in the definition of the information channel capacity \cite{Bengtsson_Zyczkowski,Jensen_Shannon properties}. \\
The method of nonparametric estimation that enables the statistical selection of the equation of motions (or generating equations) of various field theory 
or statistical physics models is  called the extremal physical information (EPI). 
The central quantity of EPI analysis is the information channel capacity  $I$, which 
is the trace of the expectation value of the $I_{F}$ matrix. 
Fundamentally, it enters into the EPI formalism as the  second order coefficient in the 
Taylor expansion of the log-likelihood function  \cite{Dziekuje informacja_2}.  
Originally, EPI was proposed  by Frie\-den and Soffer \cite{Frieden}. 
They used two Fisherian information quantities: the intrinsic information ${\rm J}$ of the source phenomenon and the information channel capacity $I$, which connects the phenomenon and observer. Both ${\rm J}$ and $I$, together with their densities, are used in the construction of two information principles, the structural and variational ones.  
${\rm J}$ and $I$ are the 
counterparts\footnote{In the sense that they are similar in relating them \cite{Jaynes}.}  
of the Boltzmann and Shannon entropies, respectively, however, they are physically different from them \cite{Frieden,Bengtsson_Zyczkowski}. 
Finally, although in the Frie\-den and Soffer approach the structural information principle is postulated on the expected level, it is then reformulated to the observed one, together with the variational principle giving two coupled differential equations. 
  \\
In \cite{Dziekuje informacja_1} two 
information principles were also postulated, but with a different interpretation of the {\it structural information}, which is denoted by $Q$. In \cite{Dziekuje informacja_2} the derivations, firstly of the observed and secondly of the expected {\it structural information principle} from basic principles was given.
It was based on the analyticity of the logarithm of the likelihood function, which allows for its Taylor expansion 
in the neighborhood of the true value of the vector parameter $\Theta$,  and on the metricity of the statistical space ${\cal S}$ of the system. 
The analytical structure of the structural information principle and the geometrical meaning of the variational information principle, which leads to the Euler-Lagrange equations, are discussed in  \cite{Dziekuje za skrypt,Dziekuje za zasadyJJ}. 
Both information principles made the EPI method the fundamental tool in the physical model selection, which is a kind of nonparametric estimation, having, as the output of the solution of the information principles the equations of motion or distribution generating equation. Their usage for the derivation of the basic field theory equations of motion is thus a fundamental one, as they anticipate these equations   \cite{Frieden,Dziekuje za skrypt,Dziekuje za models building}. 
The fact that the formalism of the information principles is used for the derivation of the distribution generating equation  
signifies that\footnote{In 
agreement with the Jaynes’ principle \cite{Jaynes}.
}  
the microcanonical description of the thermodynamic properties of a compound system has to meet the analyticity and metricity assumptions as well. \\ 
Thus, it is obvious from the former work that 
both the form of $I$ and its density  \cite{Dziekuje informacja_2,Dziekuje za skrypt,Dziekuje za zasadyJJ} play crucial roles in the construction of a particular physical model. 
Frieden, Soffer along with Plastino and Plastino \cite{Frieden} put into practice the solution of the differential information principles for various EPI models. Previously, in \cite{Frieden} the role of $I$ was discussed in many field theory contexts 
and in \cite{Amari Nagaoka book,Dziekuje informacja_2,Dziekuje za skrypt} the general view on the construction of $I$ was also presented. The main topic of the present paper will concentrate on the construction of $I$ for field theory models with the estimation performed in the base space with the Minkowski metric. \\
An important model parameter in the EPI analysis is the dimension $N$ of the sample which via the likelihood function of the system enters into the channel information capacity $I$. 
The physical models form two separate categories with respect to $N$. In the first one, to which wave mechanics and field theories belong, both $N$ and $I$ are finite \cite{Dziekuje za skrypt}. 
Classical mechanics, on the base space ${\cal Y}$ continuum,  forms the second class\footnote{In 
the contrast to e.g. wave mechanics, in classical mechanics the solution of the equation of motion does not determine (from this equation) the structure of a particle, which has to be determined independently at every point of the particle trajectory by the definition of its point structure, e.g. by the means of the $\delta$-Dirac distribution.
} 
having infinite $N$ and thus infinite $I$.
This fact was applied to prove the impossibility of the derivation of wave and field theory models from classical mechanics in \cite{Dziekuje informacja_1}. 
In the case of the first category, the sample size $N$ is the rank of the field  of the system \cite{Frieden}. 
For example, with $N=8$, the EPI method can result in the Dirac wave equation, whereas with $N=32$ it can result in the Rarita-Schwinger one \cite{Frieden}. In the realm of statistical physics, e.g. for $N=1$, the equilibrium Maxwell–Boltzmann velocity law is obtained, while for $N > 1$, the non-equilibrium, although still stationary solutions (that otherwise follow from the Boltzmann transport equation) were discovered \cite{Frieden}. 
Since the observed structural information was obtained and the new interpretation of the structural information $Q$ established \cite{Dziekuje informacja_2,Dziekuje za zasadyJJ,Dziekuje za skrypt},  some of these models have been recalculated 
\cite{Dziekuje za skrypt,Dziekuje za models building,Dziekuje za produkcyjnosc}. It appears that for every field, $N$ can be related to  the dimension of the representation of the group of symmetry transformation of the field in question \cite{Frieden}. 
\\
The paper will also deal with the kinematic form of channel information capacity $I$ expressed in terms of the point probability distributions of the sample. 
For $N=1$, the usefulness of this form is perceived in the proof of the $\mathbf{I}$--theorem  \cite{Frieden,Dziekuje za skrypt}, which is the informational analog of the $\mathbf{H}$--Boltzmann theorem. Then,  the Fisher temperature, which is the sensitivity of the Fisher information to the change of the scalar parameter, can be defined. Thus, the $\mathbf{I}$--theorem is in a sense more general than its older thermodynamic predecessor, as it can describe not only the thermodynamic properties of the compound system \cite{Jasiu i Ed temperatura}  but also of an elementary one-particle system.

\section[The basics: Rao-Fisher metric and Rao-Cram{\'e}r theorem]{The basics: Rao-Fisher metric and Rao-Cram{\'e}r theorem}

\label{alfa koneksja}

Suppose that the original random variable $Y$ takes 
vector values ${\bf y} \in {\cal Y}$ and let the vector parameter $\theta$ of 
the distribution $p({\bf y})$, in which we are interested be the {\it expected parameter}, i.e. the expectation value of $Y$: 
\begin{eqnarray}
\label{wartosc oczekiwana EY}
\theta \equiv E(Y)= \int_{\cal Y} d {\bf y}\, p({\bf y})\,  {\bf y} \; .
\end{eqnarray}
Let us now consider the $N$-dimensional sample $\widetilde{Y} = (Y_{1},Y_{2},...,Y_{N}) \equiv( Y_{n})_{n=1}^{N}$, where every $Y_{n}$ is the variable $Y$ in the $n$-th population, $n=1,2,...,N$, which is characterized by the value of the vector parameter $\theta_{n}$. 
The specific realization of $\widetilde{Y}$ takes the form
$y=({{\bf y}_{1},{\bf y}_{2},...,{\bf y}_{N}})\equiv ({\bf y}_{n})_{n=1}^{N}$, where
every datum ${\bf y}_{n}$ is generated from the distribution $p_{n}({\bf y}_{n}|\Theta)$
of the random variable $Y_{n}$, where 
the vector parameter  $\Theta$ is given by:
\begin{eqnarray}
\label{parametr Theta}
\Theta = (\theta_{1}, \theta_{2},...,\theta_{N})^{T} \equiv (\theta_{n})_{n=1}^{N} \; , \;\;\;  \\ 
\theta_{n} = (\vartheta_{1n},\vartheta_{2n},...,\vartheta_{kn})^{T} \equiv ((\vartheta_{s})_{s=1}^{k})_{n} \; . \nonumber
\end{eqnarray} 
The set of all possible realizations $y$ of the sample $\widetilde{Y}$ forms the sample space ${\cal B}$ of the system. 
%
%
In this paper, we assume that the variables $Y_{n}$ of the sample $\widetilde{Y}$ are independent. Hence, the expected parameter 
$\theta_{n'}  =  \int_{\cal B} dy \, P(y|\Theta) \, {\bf y}_{n'} $ does not influence  the {\it point probability} distribution $p_{n}({\bf y}_{n}|\theta_{n})$ for the sample index $n' \neq n$. 
The data are generated in agreement with the point probability distributions, which fulfill the condition:  
\begin{eqnarray}
\label{rozklady punktowe polozenia}
p_{n}({\bf y}_{n}|\Theta) = p_{n}({\bf y}_{n}|{\theta}_{n}) \; , \;\;\;\; {\rm where} \;\;\; n=1,...,N \; ,
\end{eqnarray}
and {\it the likelihood function} $P(y\,|\Theta)$ of the sample $y = ({\bf y}_{n})_{n=1}^{N}$ is the product:
\begin{eqnarray}
\label{funkcja wiarygodnosci proby - def}
P(\Theta) \equiv P\left({y|\Theta}\right) = \prod\limits_{n=1}^{N} {p_{n}\left({{\bf y}_{n}|\theta_{n}}\right)} \; .
\end{eqnarray}
The set of values of $\Theta=(\theta_{n})_{n=1}^{N}$ forms the coordinates of $P(y\,|\Theta)$, which is a point in $d=k \times N$ - dimensional statistical (sub)space ${\cal S}$ \cite{Amari Nagaoka book}. The number of all parameters is equal to $d=k \times N$, but for  simplicity, we will use the notation $\Theta \equiv  (\theta_{1},\theta_{2},...,\theta_{d})^{T} \equiv (\theta_{i})_{i=1}^{d}$, where the index $i=1,2,...,d$ replaces the pair of indexes  ''$sn$''. 
The likelihood function is formally the joint probability distribution of the reali\-zation $y \equiv ({\bf y}_{n})_{n=1}^{N}$ of the sample $\widetilde{Y} \equiv( Y_{n})_{n=1}^{N}$; 
hence, $P$ is the probability measure on ${\cal B}$. 
The set of all measures $\Sigma({\cal B})$ on ${\cal B}$ is {\it the state space} of the model.\\
\\
{\bf The Fisher information matrix}: Let us now examine a subset ${\cal S} \subset \Sigma({\cal B})$, on which the coordinate system $(\xi^{i})_{i=1}^{d}$  is given \cite{Amari Nagaoka book} so that the statistical space ${\cal S}$ is the $d$ - dimensional
manifold\footnote{
%
In this paper, we are interested in only the global coordinate systems.
%
}. 
Assume that ${\cal B}$ the $d$~-~dimensional statistical model:
\begin{eqnarray}
\label{model statystyczny S}
{\cal S} = \{P_{\Theta} \equiv P(y|\Theta),  \Theta \equiv (\theta_{i})_{i=1}^{d} \in  {V}_{\Theta} \subset \Re^{d} \} \; ,
\end{eqnarray}
is given, i.e. the family of the probability distributions parameterized by $d$ non-random variables
$(\theta_{i})_{i=1}^{d}$ which are real-valued and belong to the parametric space  ${V}_{\Theta}$ of the parameter $\Theta$, i.e. $\Theta \in V_{\Theta} \subset \Re^{d}$.
Thus, the logarithm of the likelihood function  $\ln  P: V_{\Theta} \rightarrow \Re$ is defined on the space $V_{\Theta}$. 
\\
Let $\tilde{\Theta} \equiv (\tilde{\theta}_{i})_{i=1}^{d} \in V_{\Theta}$ be another value of the parameter or a value of the 
estimator $\hat{\Theta}$ of the parameter $\Theta=(\theta_{i})_{i=1}^{d}$. 
At every point, $P_{\Theta}$, the $d \times d$ - dimensional observed Fisher information (FI) matrix can be defined \cite{Pawitan,Dziekuje informacja_2,Dziekuje za skrypt}:
\begin{eqnarray}
\label{observed IF}
\texttt{i\!F}(\Theta) \equiv 
- \, \partial^{i'} \partial^{i} \ln P(\Theta)   
= \left( - \, \tilde{\partial}^{i'} \tilde{\partial}^{i} \ln P(\tilde{\Theta}) \right)_{|_{\widetilde{\Theta} = \Theta}} \, 
\end{eqnarray}
and  $\partial^{i} \equiv \partial/\partial \theta_{i}$, $\tilde{\partial}^{i} \equiv \partial/\partial \tilde{\theta}_{i}$, $\,i,i'=1,2,..,d$. 
%
%
It characterizes the local properties of 
$P(y|\Theta)$. It is  symmetric and in field theory and statistical physics models with continuous, regular \cite{Pawitan}  and normalized distributions, it is positively definite. We restrict the considerations to this case only. 
The expected $d \times d$ - dimensional FI matrix on  ${\cal S}$ at point $P_{\Theta}$ \cite{Amari Nagaoka book} is defined as follows:
\begin{eqnarray}
\label{infoczekiwana}
I_F \left(\Theta\right) \equiv E_{\Theta} \left(\texttt{i\!F}(\Theta)\right) = \int_{\cal B} dy P(y|\Theta) \, \texttt{i\!F}(\Theta) \; ,
\end{eqnarray} 
where the differential element 
$dy \equiv d^{N}{\bf y} = d{\bf y}_{1} d{\bf y}_{2} ... d{\bf y}_{N}$. 
The subscript $\Theta$ in the expected value 
signifies the true value of the parameter under which the data $y$ 
are generated. 
The FI matrix defines 
on ${\cal S}$ the riemannian Rao-Fisher metric $g^{ij}$, which in the coordinates  $(\theta_{i})_{i=1}^{d}$  has the form $(g^{ij}(\Theta)) := I_F$. 
Under regularity and normalization conditions $\int_{\cal B} d^{N}{\bf y} \, P(\Theta) = 1$, \cite{Pawitan}, we obtain $\int_{\cal B} dy$ $P(\Theta)$ $\partial^{i} \ln P(\Theta) = 0\,$, $\,i=1,2,..,d$. 
Therefore, the elements $g^{ij}$ of $I_{F}$ can be rewritten as follows:  
\begin{eqnarray}
\label{Fisher inf matrix plus reg condition}
\;\;\;\; g^{ij}(\Theta) &=& -  E_{\Theta}\left(\partial^{i} \partial^{j} \ln P(y|\Theta)\right)  =  -  \int_{{\cal B}} dy \, P(y|\Theta) \, \partial^{i} \partial^{j}  \ln P(y|\Theta)  \\
&=&  \! \int_{{\cal B}} \! dy \, P(y|\Theta) \, \partial^{i} \ln P(y|\Theta) \, \partial^{j} \ln P(y|\Theta) \nonumber \\
&=& E_{\Theta}\left(\partial^{i} \ln P(y|\Theta) \, \partial^{j} \ln P(y|\Theta)\right)
\; , \;\;\; i,j = 1,2,...,d \; , \;\; \forall\, P_{\Theta} \in {\cal S} \; . \nonumber
\end{eqnarray} 
Owing to the last line, $\texttt{i\!F} = (\texttt{i\!F}^{i' i})$ 
is sometimes recorded in the "quadratic" form: 
\begin{eqnarray}
\label{observed IF Amari} 
\texttt{i\!F} = \left( \partial^{i'} \! \ln P(\Theta)\;
\partial^{i}  \ln P(\Theta) \right) \; ,
\end{eqnarray}
as it is useful in the definition of the $\alpha$-connection on the statistical space  ${\cal S}$  \cite{Amari Nagaoka book}.\\
\\
{\bf The multiparametric Cram{\'e}r-Rao theorem}: 
Let $I_{F}(\Theta)$ be the Fisher information matrix (\ref{infoczekiwana}) for $\Theta \equiv (\theta_{i})_{i=1}^{d}$ and 
$\hat{\theta}_{i}=\hat{\theta}_{i}\left({\widetilde{Y}}\right)$ be an unbiased estimator of the distinguished parameter $\theta_{i}$: 
%
%
\begin{eqnarray}
\label{ET dla DORC wielopar}
E_{\Theta} \hat{\theta}_{i} = \theta_{i} \in \Re \;
\end{eqnarray} 
and the values of the remaining parameters 
may also be unknown, i.e. they 
are to be estimated from the sample simultaneously with $\theta_{i}$.  
Then, the variance of $\hat{\theta}_{i}$ fulfills the Cram{\'e}r-Rao (CR) inequality ${\sigma^{2}_{\Theta}}\left(\hat{\theta}_{i}\right) \ge \left[{ I_{F}^{-1}\left( \Theta \right)}\right]_{ii} =:  I_{F}^{ii}\left(\Theta \right)$,  where $I_{F}^{-1}$ is the inverse of $I_{F}$ \cite{Pawitan}. 
$I_{F}^{ii}\left(\Theta \right)$ is the {\it lower bound in the Cram{\'e}r-Rao} (CRLB) inequality of the variance of the estimator $\hat{\theta}_{i}$   \cite{Pawitan}.  \\
Let $I_{F}^{i} \equiv I_{F}^{ii}{(\Theta)}$ and $I_{F i}  \equiv I_{F ii}\left(\theta_{i} \right) =  I_{F ii}{(\Theta)}$ denote the $(i,i)$ elements of the matrix $I_{F}^{-1}\left(\Theta\right)$ and $I_{F}\left(\Theta\right)$, respectively. 
%
%
In the multiparametric case discussed, the pair of inequalities proceed  \cite{Pawitan,Dziekuje za skrypt}: 
\begin{eqnarray}
\label{porownanie sigma I11 z I11do-1}
\sigma^{2}_{\Theta}\left(\hat{\theta}_{i}\right) \geq   I_{F}^{i} \geq \frac{1}{ I_{F i}} \; , \;\; {\rm where} \;\;\; 1 \leq i  \leq d \; \, ,
\end{eqnarray}
where the first one is the CR inequality. If $\Theta$ is the scalar parameter, or if the $\theta_{i}$ parameter is estimated only and the others are known, then in the second inequality in (\ref{porownanie sigma I11 z I11do-1}) the equality $I_{F}^{i} = 1/I_{F i}$ remains. In this paper, we maintain the name of the Fisher information of the parameter $\theta_{i}$ for $I_{F i}$ regardless of whether the other parameters are simultaneously estimated.

\section[The channel information capacity for the position random variable]{The channel information capacity for the position random variable}

\label{Pojecie kanalu informacyjnego}

Let the vector value ${\bf y} \in {\cal Y}$ of $Y$ be the space position vector. It could  be the space-time point  ${\bf y} \equiv ({\bf y}^{\nu})_{\nu=0}^{3}$ of the four-dimensional Minkowski  space ${\cal Y} \equiv \Re^{4}$, which occurs in 
the description of the system, e.g. in wave mechanics, or the space point ${\bf y} \equiv ({\bf y}^{\nu})_{\nu=1}^{3} \in {\cal Y} \equiv \Re^{3}$ in the three-dimensional Euclidean 
space\footnote{In 
the derivation of the equations generating the distribution in  statistical physics, ${\bf y}$ can be the value of the energy $\epsilon$ of the system  \cite{Frieden} and then  ${\bf y} \equiv \epsilon \in {\cal Y} \equiv \Re$.
}. 
Thus, ${\bf y} \equiv ({\bf y}^{\nu})$ can possess the vector index  $\nu, \mu = (0),1,2,3, ... \;$, where $\nu, \mu = 0,1,2,3, ... \;$ in the Minkowski space and $\nu, \mu = 1,2,3, ... \;$ in the Euclidean one. In this analysis, we will use random variables with covariant and contravar\-iant coordinates. The relation between them, both for the values of the random position vector and for the corresponding expectation values, is as follows:
\begin{eqnarray}
\label{wsp theta kontra i kowariantne}
{\bf y}_{\nu} = \sum_{\mu} \eta_{\nu \mu} \, {\bf y}^{\, \mu} \; , \;\;\;\;\;\; \theta_{\nu} = \sum_{\mu} \eta_{\nu \mu} \, \theta^{\, \mu} \; ,
\end{eqnarray}
where $(\eta_{\nu \mu})$ is the metric tensor of the space ${\cal Y}$. In the case of the vectorial Minkowski index, we take the following diagonal form of the metric tensor:
\begin{eqnarray}
\label{metryka M}
(\eta_{\nu \mu}) = {\rm diag}(1,-1,-1,-1,...) \;\, ,
\end{eqnarray} 
whereas for the Euclidean vectorial index, the metric tensor takes the form:
\begin{eqnarray}
\label{metryka E}
(\eta_{\nu \mu}) = {\rm diag}(1,1,1,...) \;\, .
\end{eqnarray} 
The introduction of the metric tensor $\eta$ is important from the measurement point of view.
In the measurement of the chosen $\nu$-th
coordinate of the position, we are not able to exclude the displacements (fluctuations) of the values of the space-time coordinates  which are orthogonal to it.
This indicates that the expectation value of the $\nu$-th coordinate  of the position is not calculated in (\ref{wartosc oczekiwana EY}) from a distribution of the type $p({\bf y}^{\nu})$, but that it has to be calculated from the joint distribution  $p({\bf y})$  for all coordinates  ${\bf y}^{\nu}$. In addition, the measurement which is independent of the 
coordinate system is the one of the square length $ {\bf y} \cdot {\bf y}$ 
and not of the single coordinate  ${\bf y}^{\nu}$ only, where "$\cdot$" denotes the inner product defined by the metric tensor (\ref{metryka M}) or (\ref{metryka E}). \\ 
The fact of the statistical dependence of the spacial position variables for the different indexes  ${\nu}$ should not be confused with the {\it analytical independence} which they possess. This means that the variable $Y$ is the so-called {\it Fisherian variable} for which:
\begin{eqnarray}
\label{zmienne Fisherowskie}
\partial {\bf y}^{\nu}/\partial {\bf y}^{\mu} = \delta^{\nu}_{\mu}  \; .
\end{eqnarray}
Let the data $y \equiv ({\bf y}_{n})_{n=1}^{N}$ 
be a realization of  the $N$-dimensional sample  $\widetilde{Y}$ 
for the {\it positions of the system}, 
where ${\bf y}_{n} \equiv ({\bf y}_{n}^{\nu})$, $n=1,2,...,N$, denotes the $n$-th vectorial obser\-vation. 
Now, as the number of the parameters $\theta_{n}$, where $n=1,2,...,N\,$,  agrees with the dimension $N$ of the sample, and 
as $\nu$ is the vectorial index of the coordinate 
${\bf y}_{n}^{\nu}$ (and therefore the vector parameter has the additional vectorial index $\nu$, too) thus, we have:
\begin{eqnarray}
\label{parametr wektorowy}
\Theta = \left({{\theta}_{1},{\theta}_{2},...,{\theta}_{N}}\right) \; , \;\;\; {\rm where} \;\;\;\; {\theta}_{n}=\left({\theta_{n}^{\nu}} \right) \; ,  \;\;\;\; \nu = (0),1,2,3, ... \; ,
\end{eqnarray}
where the expected parameter is the expectation value of the position of the system at the $n$-th measurement point of the  sample:
\begin{eqnarray}
\label{wartosc oczekiwana EYn}
\theta_{n} \equiv E(Y_{n}) =  (\theta_{n}^{\nu}) \; , \;\;\; {\rm where} \;\;\;\;  \theta_{n}^{\nu} = \int_{\cal B} dy \, P(y|\Theta) \, {\bf y}_{n}^{\nu}  \; .
\end{eqnarray}
Here for ${\bf y} \equiv ({\bf y}^{\nu})_{\nu=0}^{3}$ $\in {\cal Y} \equiv \Re^{4}$, $d y:= d^{4}{\bf y}_{1}...d^{4}{\bf y}_{N}$ and $d^{4}{\bf y}_{n}=d {\bf y}_{n}^{0} d {\bf y}_{n}^{1} d {\bf y}_{n}^{2} d {\bf y}_{n}^{3}$. \\
\\
{\bf The statistical spaces ${\cal S}$ and ${\cal S}_{N \times 4}$}: Let the discussed space be the Minkows\-ki space-time ${\cal Y} \equiv \Re^{4}$. Then, every one of the  distributions $p_{n}({\bf y}_{n}|\theta_{n})$ is the point of the statistical model ${\cal S} = \{p_{n}({\bf y}_{n}|\theta_{n})\}$, which is parameterized by the natural parameter, i.e. by the expectation value   $\theta_{n} \equiv (\theta_{n}^{\nu})_{\nu=0}^{3} = E(Y_{n})$, as in (\ref{wartosc oczekiwana EYn}). 
Consequently, the dimension of the sample space ${\cal B}$ and the dimension of the parametric space $V_{\Theta}$ of the vector parameter  $\Theta \equiv (\theta_{n}^{\nu})_{n=1}^{N}$ are equal to each other\footnote{Nevertheless, 
let us remember that, in general, the dimension of the vector of parameters $\Theta \equiv (\theta_{i})_{i=1}^{d}$ and the sample vector $y \equiv ({\bf y}_{n})_{n=1}^{N}$ can be different. 
} 
and, as the sample $\widetilde{Y}$ is $N \times 4$-dimensional random variable, hence the set ${\cal S}_{N \times 4}=\{p_{n}(y|\Theta)\}$ is the statistical space on which the parameters  $(\theta_{n}^{\nu})_{n=1}^{N}$ form the $N \times 4$-dimensional local coordinate system.

\subsection{The Rao-Cram{\'e}r inequality for the position random variable}

\label{Poj inform zmiennej los poloz}

According to (\ref{Fisher inf matrix plus reg condition}),  
the channel information capacity in the {\it single $n$-th} (measurement) {\it information channel}, i.e. the {\it Fisher information for the parameter} ${\theta}_{n}$ is equal to: 
\begin{eqnarray}
\label{Fisher info dla theta n gradient}
I_{F n} &\equiv& I_{F n} \left(\theta_{n} \right) = \int_{\cal B} {dy \, P\left(y|\Theta\right)} \, \left( \, \nabla_{\theta_{n}} \ln P\left(y|\Theta\right) \cdot \nabla_{\theta_{n}} \ln P\left(y|\Theta\right) \, \right) \nonumber \\
& = &
{\int_{\cal B}{dy \; P\left(y|\Theta\right) \sum\limits_{\nu, \, \mu = (0),1,2,...} \eta^{\nu \mu} {\left(\frac{\partial\ln P\left(y|\Theta\right)}{\partial\theta_{n}^{\nu}} \; \frac{\partial \ln P \left(y|\Theta\right)}{\partial\theta_{n}^{\mu}} \right)}}}
\; ,
\end{eqnarray}
where the tensor $(\eta^{\nu \mu})$
is dual to $(\eta_{\nu \mu})$, i.e. $\sum\limits_{\mu=(0),1,2,...} \!\!\!\! \eta_{\nu \mu} \eta^{\gamma \mu}$ $= \delta_{\nu}^{\gamma}$, where $\delta_{\nu}^{\gamma}$ is the Kronecker delta and
$\nabla_{\theta_{n}} = \!\!\! \sum\limits_{\mu=(0),1,2,...}  \!\! \frac{\partial}{\partial \theta_{n}^{\, \mu}} \, d {\bf y}_{n}^{\mu} \, $. \\
Simultaneously, the variance of 
the estimator $\hat{\theta}_{n} \left({y}\right)$ of the parameter $\theta_{n}$ 
has the form:
\begin{eqnarray}
\label{wariancja estymatora theta i alfa}
\;\;\;\;\;\;\; {\sigma^{2}} \,( \hat{\theta}_{n}) &=& \int_{\cal B} dy \, P\left(y|\Theta\right) \left(\hat{\theta}_{n} \left(y\right) - \theta_{n} \right) \cdot \left(\hat{\theta}_{n}  \left(y\right)-\theta_{n} \right) \\
&=& \int_{\cal B} dy \, P\left(y|\Theta\right) \sum\limits_{\nu, \, \mu = (0),1,2,...} \eta_{\nu \mu} \left(\hat{\theta}_{n}^{\nu} \left(y\right) - \theta_{n}^{\nu} \right) \, \left(\hat{\theta}_{n}^{\mu}  \left(y\right) - \theta_{n}^{\mu} \right)
\; . \nonumber
\end{eqnarray}
Let us now observe that for the variables of the space position,  
the integrals in (\ref{Fisher info dla theta n gradient}) and (\ref{wariancja estymatora theta i alfa}) are not to 
be factorized with respect to the vectorial $\nu$-th coordinate. 

Now, as a consequence of (\ref{rozklady punktowe polozenia}),
for every distinguished parameter  $\theta_{n}\,$, the variance  ${\sigma^{2}} \, (\hat{\theta}_{n})$ of its estimator given by (\ref{wariancja estymatora theta i alfa}) is connected with the Fisher information  $I_{F n} = I_{F n}(\theta_{n})$  given by (\ref{Fisher info dla theta n gradient}) 
in the single 
information channel for this parameter 
by the inequality (\ref{porownanie sigma I11 z I11do-1}): 
\begin{eqnarray}
\label{Rao Cram uogolnienie}
\frac{1}{{\sigma^{2}}\,\left(\hat{\theta}_{n} \right)} \leq  \frac{1}{I_{F}^{n}} \leq  I_{F n} \;\;\; {\rm where} \;\;\; n=1,2,...,N \; ,
\end{eqnarray}
where $I_{F}^{n}$ is the CRLB  for the parameter ${\theta}_{n}$.
\\
\\
{\bf The Stam's information}:  The quantity $\frac{1}{{\sigma^{2}} \, (\hat{\theta}_{n})}$ refers 
to the single $n$-th channel.  The {\it Stam's information} $I_{S}$ \cite{Stam,Frieden} is obtained by summing over 
its index $n$ and is equal to:
\begin{eqnarray}
\label{informacja Stama}
0 \leq  I_{S} \equiv \sum\limits_{n=1}^{N} \frac{1}{{\sigma^{2}}\,(\hat{\theta}_{n})} =: \sum\limits_{n=1}^{N} I_{S n} \;\, ,
\end{eqnarray}
where ${\sigma^{2}} \,( \hat{\theta}_{n})$ is given in (\ref{wariancja estymatora theta i alfa}). 
This is the {\it scalar} measure of the quality of the simultaneous estimation in all information channels. {\it The Stam's information is by definition always nonnegative}.
As $\theta_{n} = (\theta_{n}^{\nu})$ is the vectorial parameter, $I_{S n} = \frac{1}{{\sigma^{2}}\,(\hat{\theta}_{n})}$ itself is the Stam's information of the (time-) space
channels for the $n$-th measurement in the 
sample\footnote{The 
appearance of the Minkowski metric in the 
Stam's information $I_{S n}$ in (\ref{informacja Stama}) and (\ref{wariancja estymatora theta i alfa}) justifies the 
use of the error propagation law (and the calculation of the mean square of the measurable quantity) in the arbitrary Euclidean metric, that is, when in addition to the spacial indexes  $x_{k}$, $k=1,2,3,...$,  the temporal index $t$ occurs with the imaginary unit $i$ 
\cite{Sakurai 2,Frieden}. 
}. 
\\ 
Finally, summing the LHS and RHS of (\ref{Rao Cram uogolnienie})
over the index $n$ 
and  taking into account (\ref{informacja Stama}),
we observe that  $I_{S}$ fulfills the inequality: 
\begin{eqnarray}
\label{informacja Stama vs pojemnosc informacyjna}
0 \leq I_{S} \equiv \sum\limits_{n=1}^{N} I_{S n} \le \sum\limits_{n=1}^{N} {I_{F n}} =: I 
\; ,
\end{eqnarray}
where $I$, denoted in the statistical literature by $C$, is the {\it  channel information capacity}   \cite{Bengtsson_Zyczkowski,Jensen_Shannon properties} 
{\it of the system}\footnote{
Thus the channel information capacity has the form: 
\begin{eqnarray}
\label{I dla niezaleznych Yn}
I =  \sum\limits_{n=1}^{N} {I_{F n}} = \sum\limits_{n=1}^{N}  \int_{\cal B} \! d y P(y|\Theta) \, \sum_{\nu=0}^{3} \texttt{i\!F}_{n n \nu}^{\;\;\;\;\, \nu}  = \int_{\cal B} d y\, P(y|\Theta) \; \sum_{n=1}^{N} \texttt{i\!F}_{n n} = \int_{\cal B} d y\, \textit{i}  \; , 
\end{eqnarray}
where $\textit{i} :=  P(\Theta) \; \sum_{n=1}^{N} \texttt{i\!F}_{n n}$ is the {\it channel information density}   \cite{Dziekuje informacja_2,Dziekuje za skrypt,Dziekuje za zasadyJJ}. \\
The channel information capacity $I_{Fn}$ is invariant under the smooth  invertible mapping $Y \rightarrow  X$, where $X$ is the new variable \cite{Streater}.
It is also invariant under the space and time reflections. 
}: 
\begin{eqnarray}
\label{pojemnosc C}
I = \sum\limits_{n=1}^{N}{I_{F n}} \; .
\end{eqnarray}
The inequality (\ref{informacja Stama vs pojemnosc informacyjna}) is the 'minimal', {\it trace } generalization of the "single-channel" CR inequality (\ref{Rao Cram uogolnienie}). 
\\
From the physical modeling point of view, the 
channel information capacity  $I$ is the most important statistical concept,  which lays the foundation for  
the kinematical terms \cite{Frieden} of various field theory models. 
According to (\ref{informacja Stama}), it appears that both for the Euclidean (\ref{metryka E}) and Minkowskian metric (\ref{metryka M}) we perform the estimation in the case of positive  $I_{S}$ only.   
Hence, from (\ref{informacja Stama vs pojemnosc informacyjna}) it follows that $I$ is  also non-negative. In (\ref{m E p})  Section~\ref{Informacja Fouriera}, it will be shown that  $I$ 
is non-negatively defined for the field theory with particles which have a non-negative square of their masses \cite{dziekuje za neutron,Frieden}.   
This fact for the Minkowskian space ought to be checked in any particular field theory model, but from the estimation theory it is evident that:
\begin{eqnarray}
\label{przyczynowosc}
{\sigma^{2}}\,(\hat{\theta}_{n}) \geq 0 \; ,
\end{eqnarray}
which is always the case for {\it causal processes}. \\ 
\\
{\bf Remark~1}. {\it The index of the measurement channel}: 
The sample index $n$ is the smallest index of the information channel  in which the measurement  is performed. 
It indicates that if the sample index might be additionally indexed, 
e.g. by the space-time index $\nu$, then 
it would be impossible to perform the measurement in one of the subchannels  $\nu$ assigned in that manner (without performing it simultaneously in the remaining  subchannels which also possess the sample index $n$).  The channel which is inseparable from the experimental  point of view  will be referred to as the {\it measurement channel.}  \\
{\bf Remark~2}.  
In case of the restriction of the analysis to only a part of the measurement channel, one should ascertain that the value of the Stam's information which is the subject of the analysis is positive. For example, in the case of the neglect of the temporally indexed part of the space-time measurement channel, the Stam's inequality obtained for the spacial components has the form:
\begin{eqnarray}
\label{nierownosc informacja Stama dla podkanalow przestrzennych}
\!\!\!\!\!\!\! 0 \! & \leq &  \! I_{S} =  \sum_{n=1}^{N} I_{S n} \nonumber \\
\!\!\!\!\!\!\! & \leq & \sum_{n=1}^{N} {\int{d\vec{y} \, P\left(\vec{y}|\vec{\Theta}\right) \sum \limits_{i=1}^{3}{\frac{\partial\ln P\left(\vec{y}|\vec{\Theta}\right)}{\partial\theta_{n i}} \; \frac{\partial\ln P\left(\vec{y}|\vec{\Theta}\right)}{\partial\theta_{n i}}}}} =: \sum_{n=1}^{N} I_{F n} = I\; ,
\end{eqnarray}
where the vector denotation means that the analysis of both in the sample space and in the parameter space has been 
reduced to the spacial part of the random variables and parameters, respectively. \\ 
{\bf Remark~3}. {\it The symmetry conformity of $I_{S}$ and $I$}: 
The error of the estimation of the expected 
position four-vector in the $n$-th measurement channel, and thus $I_{S n} = 1/{\sigma^{2}}\,(\hat{\theta}_{n})$, must be independent of the coordinate system in the Minkowski (or Euclidean) space. Therefore, $I_{S n}$ as defined in (\ref{informacja Stama}) and (\ref{wariancja estymatora theta i alfa}) is invariant under the Lorentz transformations (i.e. busts and rotations) for the metric tensor given by (\ref{metryka M}) or for metric tensor given by (\ref{metryka E}) under the Galilean ones. \\
As the measurements in the sample are independent, the channel information capacity  $I$ is  invariant under the Lorentz transformation in the space with the Minkowskian metric, or under the Galilean  tran\-sfor\-mation in the Euclidean space, only if every $I_{Fn}$ is invariant. 
The conditions of the invariance of $I_{S n}$ and $I_{Fn}$ converge if in the 
inequalities given by (\ref{Rao Cram uogolnienie}), the equalities are attained. More information on the invariance of the CRLB under the displacement, 
space reflection, rotation and affine transformation  as well as unitary transformation can be found in \cite{PPSV}. \\ 
{\bf Remark~4}. {\it Minimization of $I$ with respect to $N$}: Every $n$-th term, $n=1,...,N$, in the sum 
(\ref{pojemnosc C})
brings, as the degree of freedom for $I$, its analytical  contribution. If only the added degrees of freedom do not have an effect on the already existing ones, then because every $I_{Fn}$ 
is non-negative,  the channel information capacity $I$ has an increasing 
tendency with an increase of $N$. The minimization criterion of $I$ with respect to $N$ was used by Frieden and Soffer as an additional condition in the selection 
of the equation of motion for the field theory model or the generation equation in the statistical physics realm \cite{Frieden}. 
This means that values of $N$ above the minimal allowable one lack a  physical meaning for the particular application and leaving them in the theory makes it unnecessarily complex.  Yet, sometimes part of the consecutive values of $N$ are also necessary. Some examples were given in the Introduction.

\section[The kinematic information in the Frieden-Soffer approach]{The kinematic information in the Frieden-Soffer approach}

\label{Podstawowe zalozenie Friedena-Soffera}

{\bf The basic physical assumption of the  Frieden-Soffer approach}.   Let the data  $y \equiv ({\bf y}_{n})_{n=1}^{N}$ 
be the realization of the sample for the position of the system where ${\bf y}_{n} \equiv ({\bf y}_{n}^{\nu})_{\nu=0}^{3} \in {\cal Y} \equiv \Re^{4}$. {\it In accordance with the assumption proposed by  Frieden and Soffer} \cite{Frieden}, {\it their collection is carried  out 
by the system alone in accordance with the probability density distributions, $p_{n}({\bf y}_{n}|\theta_{n})$, $n=1,...,N$}.  \\
The content of the above assumption can be expressed as follows:  {\it The system {\bf samples the space-time} which is accessible to it  ``collecting the data and performing 
the statistical analysis'', in accordance with some information principles} (see \cite{Frieden,Dziekuje informacja_1,Dziekuje informacja_2,Dziekuje za skrypt,Dziekuje za zasadyJJ}). \\
{\bf Note}: 
The estimation is performed by a human and the EPI method should be perceived as only a {\it type of statistical analysis}. \\ 
\\
The statistical procedure in which we are interested concerns the inference about $p_{n}({\bf y}_{n}|\theta_{n})$ on the basis of the data $y$ using the likelihood function  $P(y|\Theta)$. 
Therefore, using (\ref{funkcja wiarygodnosci proby - def}),  
we can record the derivatives of $\ln P$ standing in $I_{F n}$ in (\ref{Fisher info dla theta n gradient}) as follows:
\begin{eqnarray}
\frac{\partial \ln P \left(y|\Theta \right)}{\partial \theta_{n\nu}} = \frac{\partial}{\partial \theta_{n\nu}} \sum\limits_{n=1}^N {\ln p_{n} \left({\bf y}_n|\theta_n \right)} = \sum\limits_{n=1}^N {\frac{1}{p_n \left({\bf y}_n|\theta_n \right)} \frac{\partial{p_n \left({\bf y}_n|\theta_n \right)} }{\partial \theta_{n\nu}}} \;
\end{eqnarray}
and applying the normalization $\int_{\cal Y}d^{4}{\bf y_{n}} p_n \left({\bf y}_n|\theta_n \right) =  1$ of each of the marginal distributions we obtain the following form of the channel information capacity:
\begin{eqnarray}
\label{postac I bez log p po theta}
I = \sum_{n=1}^N {\int_{\cal Y} d^{4}{\bf y}_n \frac{1}{{p_{n} \left(  {\bf y}_n|\theta_{n}  \right)}} \sum\limits_{\nu, \, \mu =0}^{3} \eta^{\nu \mu}  {\left( {\frac{{\partial p_{n} \left(  {\bf y}_n|\theta_{n}  \right)}}{{\partial \theta_{n}^{ \nu} }}}  {\frac{{\partial p_{n} \left(  {\bf y}_n|\theta_{n}  \right)}}{{\partial \theta_{n}^{ \mu} }}} \right) } } \; .
\end{eqnarray}
\\
Finally, as the amplitudes  $q_{n}\left( {\bf y}_n|\theta_{n}  \right)$ of the measurement data  ${\bf y}_{n} \in {\cal Y}$  are determined as in \cite{Frieden,Amari Nagaoka book,Bengtsson_Zyczkowski}: 
\begin{eqnarray}
p_{n} \left(  {\bf y}_n|\theta_{n}  \right) =: q_{n}^{2}\left(  {\bf y}_n|\theta_{n}  \right) \; ,
\end{eqnarray}
simple calculations give: 
\begin{eqnarray}
\label{potrz}
I = 4 \sum\limits_{n=1}^N \int_{\cal Y} {d^{4}{\bf y}_{n} \sum\limits_{\nu, \,\mu = 0}^{3} \eta^{\nu \mu} {\left( {\frac{{\partial q_{n} \left(  {\bf y}_n|\theta_{n}  \right)}}{{\partial \theta_{n}^{ \nu} }}}  {\frac{{\partial q_{n} \left(  {\bf y}_n|\theta_{n}  \right)}}{{\partial \theta_{n}^{ \mu} }}} \right) } } \; ,
\end{eqnarray}
which is almost the key form of the channel information capacity of the Frieden-Soffer EPI method. 
It becomes obvious that by the construction the rank of the field of the system, which is the number of the amplitudes  $\left(q_{n}(  {\bf y}_n|\theta_{n}) \right)_{n=1}^{N}$, is equal to the dimension $N$ of the sample.

\subsection[The kinematic form of the Fisher information]{The kinematic form of the Fisher information}

\label{The kinematical form of the Fisher information}

The estimation of physical equations of motion using the EPI method \cite{Frieden,Dziekuje informacja_2,Dziekuje za skrypt,Dziekuje za zasadyJJ} is often connected with the necessity of rewriting $I$, originally defined on the statistical space  ${\cal S}$, in a form which uses the  displacements defined on the base space ${\cal B}$ of the sample. The task is performed as follows:  
Let  ${\bf x}_{n} \equiv({\bf x}^{\nu}_{n})$ be the displacements (e.g. the additive fluctuations) of the data ${\bf y}_{n} \equiv ({\bf y}_{n}^{\nu})$ from their expectation values $\theta^{\nu}_{n}$, i.e.:
\begin{eqnarray}
\label{parameters separation}
{\bf y}_{n}^{\nu} = \theta^{\nu}_{n} + {\bf x}^{\nu}_{n} \; .
\end{eqnarray}
In accordance with (\ref{zmienne Fisherowskie}), the displacements ${\bf x}^{\nu}_{n}$ are the Fisherian  variables. \\
Appealing to  the ``chain rule'' for the derivative \cite{Frieden}:
\begin{eqnarray}
\label{chain rule}
\frac{\partial}{\partial {\bf \theta_{n}^{\nu}}} =  \frac{\partial ({\bf y_{n}^{\nu}} -
\theta_{n}^{\nu})}{\partial \theta_{n}^{\nu}} \, \frac{\partial}{\partial ({\bf
y_{n}^{\nu}} - \theta_{n}^{\nu})}  = - \; \frac{\partial}{\partial
({\bf y_{n}^{\nu}} - \theta_{n}^{\nu})} = - \; \frac{\partial}{\partial {\bf x_{n}^{\nu}}} \; ,
\end{eqnarray}
and taking into account $d^{4}{\bf x}_{n}=d^{4}{\bf y}_{n}$, 
we can switch from the statistical form (\ref{potrz})
to the {\it kinematic form} of the channel information capacity:
\begin{eqnarray}
\label{Fisher_information-kinetic form}
I  =  4  \sum_{n=1}^{N} \int_{{\cal X}_{n}} \!\! d^{4}{\bf x}_{n}  \sum\limits_{\nu, \,\mu = 0}^{3} \eta^{\nu \mu} \frac{\partial q_{n}({\bf x}_{n}+\theta_{n}|\theta_{n})}{\partial {\bf x}_{n}^{ \nu}}
\frac{\partial q_{n}({\bf x}_{n}+\theta_{n}|\theta_{n})}{\partial {\bf x}_{n}^{\mu}} \; ,
\end{eqnarray}
where 
${\cal X}_{n}$ is the space of the ${\bf x}_{n}$ displacements. 
In rewriting (\ref{potrz}) in the form of (\ref{Fisher_information-kinetic form}), the equality:
\begin{eqnarray}
\label{zapis dla qn w xn}
q_{n}({\bf y}_{n}|\theta_{n}) = 
q_{n}({\bf x}_{n}+\theta_{n}|\theta_{n}) 
 \; 
\end{eqnarray}
has been used. \\
Assuming that the range of the variability of all  ${\bf x}_{n}^{\nu}$ is the same for every  $n$, we can disregard 
the index $n$ for these variables (but not for the amplitudes  $q_{n}$) and thus obtain the formula: 
\begin{eqnarray}
\label{Fisher_information-kinetic form bez n pelny zapis}
I  =  4  \sum_{n=1}^{N} \int_{\cal X} \!\! d^{4}{\bf x}  \sum\limits_{\nu, \,\mu = 0}^{3} \eta^{\nu \mu} \frac{\partial q_{n}({\bf x} + \theta_{n}|\theta_{n})}{\partial {\bf x}^{\nu}}
\frac{\partial q_{n}({\bf x} + \theta_{n}|\theta_{n})}{\partial {\bf x}^{\mu}} \; ,
\end{eqnarray}
where ${\cal X}$ is the space of displacements ${\bf x}$. 
The resulting formula (\ref{Fisher_information-kinetic form bez n pelny zapis}) indicates that the Fisher-Rao metric (\ref{Fisher inf matrix plus reg condition}) on the statistical space ${\cal S}$ generates the {\it kinetic energy metric}. Let us now note that the kinetic term for every particular amplitude $q_{n}$ enters with different $\theta_{n}$, in  principle. Hence, in fact $N$ kinetic terms have been obtained, one for every amplitude $q_{n}({\bf x} + \theta_{n}|\theta_{n})$. \\ 
The EPI model from which, e.g. a particular field theory model appears,  is usually\footnote{The 
analysis of the EPR-Bohm experiment takes place on a parameter space $V_{\Theta}$ \cite{Frieden,Dziekuje za skrypt}.
}
built over the displacements space ${\cal X}$, which in our case is the Minkowski space-time $ R^{4}$.  
Although the statistical model  ${\cal S}$ is transformed from being defined on the parameter space $V_{\Theta} \equiv \Re^{4}$ to the space of displacements ${\cal X} \equiv  \Re^{4}$, the original sample space ${\cal B}$ remains its base space both before and after this redefinition. \\
In this way, the basic tool for the EPI model estimation connected with the derivation of both the generating equations of statistical physics \cite{Frieden,Dziekuje za produkcyjnosc} and the equations of motion  for field theory models \cite{Frieden,Dziekuje za skrypt,Dziekuje za models building}, e.g. the Maxwell electrodynamics \cite{Frieden}, is  obtained. \\
A simplifying notation will now be introduced: 
\begin{eqnarray}
\label{denotation dla qn w xn}
q_{n}({\bf x}) \equiv q_{\theta_{n}}({\bf x}) \equiv  q_{n}({\bf x} + \theta_{n}|\theta_{n}) \; ,
\end{eqnarray}
which {\it leaves the whole information on  $\theta_{n}$ that characterizes $q_{n}({\bf x} + \theta_{n}|\theta_{n})$ in the index $n$ of the amplitude $q_{n}({\bf x})$} (and similarly for the original distribution  $p_{n}({\bf x})$). With this simplifying notation, the formula (\ref{Fisher_information-kinetic form bez n pelny zapis}) can be rewritten as follows:
\begin{eqnarray}
\label{Fisher_information-kinetic form bez n}
I  =  4  \sum_{n=1}^{N} \int_{\cal X} \!\! d^{4}{\bf x}  \sum\limits_{\nu, \,\mu = 0}^{3} \eta^{\nu \mu} \frac{\partial q_{n}({\bf x})}{\partial {\bf x}^{\nu}}
\frac{\partial q_{n}({\bf x})}{\partial {\bf x}^{\mu}} \; ,
\end{eqnarray}
which is understood in accordance with  (\ref{Fisher_information-kinetic form bez n pelny zapis}).\\
\\
{\bf Note on the shift invariance}:  
In the above derivation, the assumption used in \cite{Frieden} on the {\it invariance of the distribution under the shift} ({\it displacement}): 
$ p_{n} ({\bf x}_{n}) \equiv p_{x_{n}} ({\bf x}_{n}|\theta_{n})$ $= p_{n} ({\bf x}_{n} + \theta_{n}|\theta_{n})$ $= p_{n} ({\bf y}_{n}|\theta_{n})$ , where ${\bf x}_{n}^{\nu} \equiv {\bf y}_{n}^{\nu} - \theta_{n}^{\nu}$, 
has not been used.

\subsection{The probability form of the kinematic channel information capacity}

\label{Postac kinematyczna pojemnosci zapisana w prawdopodobienstwie}

Here, the starting point is the form of $I$ given by (\ref{postac I bez log p po theta}).
We have to move on to the additive displacements
${\bf x}_{n} \equiv({\bf x}^{\nu}_{n})$ (\ref{parameters separation})  and use the ``chain rule'' 
(\ref{chain rule}). 
As was mentioned previously, 
assuming that the range of the variability of all ${\bf x}_{n}^{\nu}$ is the same for every $n$ and disregarding the variables index $n$,  
but leaving the information about  $\theta_{n}$ in the subscript of the point distributions  $p_{n}$, we obtain 
the 
{\it kinematic form} of the channel information capacity expressed as the functional of the probabilities: 
\begin{eqnarray}
\label{postac I dla p po x bez n}
I = \sum_{n=1}^N {\int_{\cal X} d^{4}{\bf x} \frac{1}{{p_{n} \left(  {\bf x}  \right)}} \sum\limits_{\nu, \,\mu = 0}^{3}  \eta^{\nu \mu}  {\left( {\frac{{\partial p_{n} \left(  {\bf x}  \right)}}{{\partial {\bf x}^{\nu} }}}  {\frac{{\partial p_{n} \left(  {\bf x}  \right)}}{{\partial {\bf x}^{ \mu} }}} \right) } } \; ,
\end{eqnarray}
where the simplifying notation $p_{n}({\bf x}) \equiv p_{\theta_{n}}({\bf x}) \equiv  p_{n}({\bf x} + \theta_{n}|\theta_{n})$,  with the same meaning as  (\ref{denotation dla qn w xn}) in (\ref{Fisher_information-kinetic form bez n}) has been used. 
The above form is used as the primary one in the construction of $I$ for Maxwell electrodynamics (see below)  
and in the weak field limit of the gravitation theory \cite{Dziekuje za skrypt}.

\subsection[The channel information capacity for Maxwell electrodynamics]{The channel information capacity for Maxwell electrodynamics}

\label{Maxwell field}

The Frieden-Soffer EPI method of the derivation of the Maxwell  equations of electrodynamics was presented in \cite{Frieden}. However, in \cite{Frieden} the form of $I$ is the Euclidean one. 
What was lacking for the construction of the Minkowski form of $I$ is the notion of the measurement channel presented in previous sections from which the fully relativistic covariant form of the channel information capacity required for the Maxwell model follows  \cite{Dziekuje za skrypt}. 
Below, we present its construction from which the EPI method for the Maxwell equations gains its full Fisherian statistical validity.

We begin with the channel information capacity written in the basic form given by (\ref{postac I dla p po x bez n}). The proof that in order to obtain the Maxwell equations of motion, the field of the rank equal to $N=4$ with real amplitudes $q_{n}$, $n=1,2,3,4$ has to be analyzed, is given in \cite{Frieden}, where one of the assumptions is that the gauge fields are  proportional to these amplitudes:
\begin{eqnarray}
\label{amplitudy dla Maxwella}
q_{\nu}({\bf x})=a\, A_{\nu}({\bf x}) \; , \;\;\; {\rm where} \;\;\; \nu \equiv n-1 = 0,1,2,3 \; , 
\end{eqnarray}
where $a$ is a constant and the Minkowski sample index $\nu$ is introduced.   
\\
Using the Minkowski metric $(\eta^{\nu\mu})$, we now define the amplitudes  $q^{\nu}({\bf x})$ which are {\it dual} to  $q_{\nu}({\bf x})$:
\begin{eqnarray}
\label{amplitudy dla Maxwella dual}
q^{\nu}({\bf x}) &\equiv& \sum_{\mu=0}^{3} \eta^{\nu \mu} q_{\mu}({\bf x}) = a\, \sum_{\mu=0}^{3}  \eta^{\nu \mu} A_{\mu}({\bf x}) \\ &\equiv& a\,A^{\nu}({\bf x}) \; , \;\;\; {\rm where} \;\;\; \nu \equiv n-1 = 0,1,2,3 \; , \nonumber
\end{eqnarray}
where the dual gauge fields $A^{\mu}({\bf x})$ have been introduced: \begin{eqnarray}
\label{pola cech dualne dla Maxwella}
A^{\nu}({\bf x}) \equiv \sum_{\mu=0}^{3}  \eta^{\nu \mu} A_{\mu}({\bf x})  \; , \;\;\; {\rm where} \;\;\; \nu  = 0,1,2,3 \; .
\end{eqnarray}
The amplitudes $q_{\nu}({\bf x})$ are connected with the point probability  distributions  $p_{n} \left({\bf x}  \right)$ in the following way:
\begin{eqnarray}
\label{rozklad p n dla Maxwella}
p_{n} \left({\bf x}  \right) &\equiv& p_{q_{\nu}} \left(  {\bf x}  \right)  =  q_{\nu}({\bf x}) q_{\nu}({\bf x}) \\
&=& a^{2} A_{\nu}({\bf x}) A_{\nu}({\bf x}) \; , \;\;\; {\rm where} \;\;\; \nu \equiv n-1 = 0,1,2,3 \; . \nonumber
\end{eqnarray}
Consequently, we observe that in the EPI method applied to Maxwell electrodynamics, the sample index $n$ becomes the space-time one. Thus, the form of the channel information capacity has to take into account the additional estimation in the space-time channels taking, in accordance with the general  prescriptions of Section~\ref{Poj inform zmiennej los poloz}, the covariant form:
\begin{eqnarray}
\label{postac I dla p po x  suma po Mink Maxwell}
I &=& \sum_{\mu=0}^{3} {\int_{\cal X} d^{4}{\bf x}\, \eta^{\mu\mu}\frac{1}{{p_{q_{\mu}} \left(  {\bf x}  \right)}} \sum\limits_{\nu=0}^{3}  {\left( {\frac{{\partial p_{q_{\mu}} \left(  {\bf x}  \right)}}{{\partial {\bf x}_{\nu} }}}  {\frac{{\partial p_{q_{\mu}} \left(  {\bf x}  \right)}}{{\partial {\bf x}^{ \nu} }}} \right) } }  \\
&=&  4 \sum_{\mu=0}^{3} {\int_{\cal X} d^{4}{\bf x}\,  \sum\limits_{\nu=0}^{3}  {\left( {\frac{{\partial q_{\mu} \left(  {\bf x}  \right)}}{{\partial {\bf x}_{\nu} }}}  {\frac{{\partial q^{\mu} \left(  {\bf x}  \right)}}{{\partial {\bf x}^{ \nu} }}} \right) } }  \; . \nonumber
\end{eqnarray}
Finally, in accordance with (\ref{postac I dla p po x  suma po Mink Maxwell}) and  (\ref{amplitudy dla Maxwella dual}), the channel information capacity  $I$ is as follows:
\begin{eqnarray}
\label{IF for Maxwell}
I =  4\, a^{2} \sum_{\mu=0}^{3} \int_{\cal X} \! d^{4}{\bf x}\,  \sum\limits_{\nu=0}^{3}  {\left( {\frac{{\partial A_{\mu} \left(  {\bf x}  \right)}}{{\partial {\bf x}_{\nu} }}}  {\frac{{\partial A^{\mu} \left(  {\bf x}  \right)}}{{\partial {\bf x}^{ \nu} }}} \right) }  \, .
\end{eqnarray}
It must be stressed that the proportionality  $q_{\nu}({\bf x}) = a A_{\nu}({\bf x})$ and the normalization condition:
\begin{eqnarray}
\label{A normalization}
(1/4)\sum_{\nu=0}^{3}\int_{\cal X} d^{4}{\bf x}\, q_{\nu}^{2}({\bf x})=\sum_{\nu=0}^{3}\int_{\cal X} d^{4}{\bf x}\, A_{\nu}^{2}({\bf x}) = 1 \; ,
\end{eqnarray}
where $a=2$, pose a question on the meaning of the localization of the photon and the existence of its wave function. 
These points have recently been discussed in the literature \cite{Roychoudhuri}. For example, the discussion  in \cite{Roychoudhuri} supports the view that places the Maxwell equations on the same footing as the Dirac one;  
a fact previously noted by Sakurai \cite{Sakurai 2}. 
It is worth noting that the normalization 
condition\footnote{The normalization (\ref{A normalization}) 
of the four-potential  $A_{\nu}$ to unity might not necessarily occur. The indispensable 
condition for $q_{\nu}({\bf x})$  is that the necessary means can be calculated  \cite{Leonhardt}. 
}
(\ref{A normalization}) of the four-potential $A_{\nu}$ harmonizes the value of the proportionality constant
$a=2$ with the rank $N=4$ of the light field whose value of $a$ also occurred previously in \cite{Frieden} but as the result of harmonizing the EPI result\footnote{Which is the result of solving \cite{Frieden} the structural and variational information principles \cite{Frieden,Dziekuje informacja_2,Dziekuje za zasadyJJ} together with $\partial_{\mu} A^{\mu}=0$, which is the Lorentz one.
}
with the Maxwell equations.\\
\\
Now, both the variational and structural (see Introduction and \cite{Frieden,Dziekuje za skrypt,Dziekuje za models building}) information principles  have to be self-con\-sis\-tently  solved with the Lorentz condition \cite{Frieden}: 
\begin{eqnarray}
\label{Lorentz for Maxwell} 
\sum_{\nu,\, \mu=0}^{3} \eta^{\mu \,\nu} \partial_{\mu} A_{ \nu} = 0 
\end{eqnarray}
additionally imposed. This task, with the proper form of the structural information for the Maxwell equations additionally found, is presented  in \cite{Frieden}. In that work, the Maxwell equations were obtained by solving the EPI information principles and a discussion of the solutions for the gauge fields can be also found there. What has been lacking is the construction of the Minkowski form of the channel information capacity, which was presented above.

\section{The Fourier information}

\label{Informacja Fouriera}

Let us now consider the particle as a system described by the field of the rank  $N$, which is the set\footnote{For the case of complex wave functions see \cite{Frieden}.
}
of the amplitudes  $q_{n}({\bf x})$, $n=1,2,...,N$ determined in
the position space-time ${\cal X}$ of displacements ${\bf x} \equiv ({\bf x}^{\mu})_{\mu=0}^{\, 3} = (c t,\, {\bf x}^{1},{\bf x}^{2},{\bf x}^{3})$ as in Section~\ref{The kinematical form of the Fisher information} 
that possesses the channel information capacity   (\ref{Fisher_information-kinetic form bez n}). 
The complex Fourier transforms $\tilde{q}_{n}({\bf p})$ of the real functions $q_{n}({\bf x})$, where ${\bf p} \equiv (\wp^{\mu})_{\mu=0}^{\,3} = (\frac{E}{c},\, \wp^{1},\wp^{2},\wp^{3})$ is the four-momentum which belongs to the energy-momentum space ${\cal P}$ conjugated to the position space-time ${\cal X}$, have the form: 
\begin{eqnarray}
\label{Fourier transf}
\tilde{q}_{n}({\bf p}) = \frac{1}{(2\pi\hbar)^{2}} \int_{\cal X} d^{4} {\bf x} \; q_{n}({\bf x}) \, e^{i\,(\,\sum_{\nu=0}^{3} {\bf x}^{\nu} \wp_{\nu})/\hbar} \; ,
\end{eqnarray}
where $\sum_{\nu=0}^{3} {\bf x}^{\nu} \wp_{\nu} = E t - \sum_{l=1}^{3} {\bf x}^{l} \wp^{l}$ and $\hbar$ is the Planck constant. \\
The Fourier transform is the {\it unitary transformation which preserves the measure} on the $L^{2}$ space  of functions integrable with the square, i.e.:
\begin{eqnarray}
\label{miara zachowana}
\int_{\cal X} d^{4}{\bf x}\,q_{n}^{*}({\bf x})\,q_{m}({\bf x}) = \int_{\cal P} d^{4}{\bf p}\,\tilde{q}_{n}^{\,*}({\bf p})\,\tilde{q}_{m}({\bf p})  \; .
\end{eqnarray}
Hence, applying the condition of the probability
normalization\footnote{
{\bf Note on the displacement probability distribution in the system}:
Appealing to  the theorem of the total probability, the probability {\it density} distribution of the displacement (or fluctuation)  in the system can be written as follows  \cite{Frieden}:
\begin{eqnarray}
\label{p jako suma po qn2 przez N}
\;\;\;  p\left({\bf x}\right) = \sum_{n=1}^{N} p\left({\bf x}|{\theta}_{n} \right) r\left({\theta}_{n}\right) = \sum_{n=1}^{N} {p_{n}\left( {\bf x}_{n}|{\theta}_{n}\right) r\left({\theta}_{n}\right)}
= \frac{1}{N} \sum_{n=1}^{N} q_{n}^{2} \left({\bf x}_{n}|{\theta}_{n}\right)
= \frac{1}{N} \sum_{n=1}^{N} q_{n}^{2} \left({\bf x}\right) \; . \nonumber
\end{eqnarray}
Function $r\left({\theta}_{n}\right) = \frac{1}{N}$ can be referred to as the ``ignorance'' function as its form is a reflection of the total lack of knowledge on which out from $N$ possible values of ${\theta}_{n}$ appears in a specific $n$-th experiment in an $N$-dimensional sample. 
}:
\begin{eqnarray}
\label{norm condition dla q}
\frac{1}{N}\sum_{n=1}^{N} \int_{\cal X} d^{4}{\bf x} \, q_{n}^{2}({\bf x}) = 1 \; ,
\end{eqnarray}
we obtain (as the consequence of the Parseval's theorem): 
\begin{eqnarray}
\label{norm condition}
\frac{1}{N}\sum_{n=1}^{N} \int_{\cal X} d^{4}{\bf x} \, q_{n}^{2}({\bf x}) = \frac{1}{N} \sum_{n=1}^{N} \int_{\cal P} d^{4} {\bf p}\,| \tilde{q}_{n}({\bf p})|^{2} = 1 \; ,
\end{eqnarray}
where $|q_{n}({\bf x})|^{2} \equiv q_{n}^{2}({\bf x})$ and $| \tilde{q}_{n}({\bf p})|^{2} \equiv \tilde{q}_{n}^{*}({\bf p})\,\tilde{q}_{n}({\bf p})$. \\
Using (\ref{Fourier transf}) we can now record $I$ given by (\ref{Fisher_information-kinetic form bez n}) as follows:
\begin{eqnarray}
\label{I by Fourier tr}
I\left[q({\bf x})\right] = I\left[\tilde{q}({\bf p})\right] = \frac{4}{\hbar^{2}} \int_{\cal P} d^{4} {\bf p} \sum_{n=1}^{N} |\tilde{q}_{n}({\bf p})|^{2}\,(\frac{E^{2}}{c^{2}} - \vec{\wp}^{\,2}\,) \, ,
\end{eqnarray}
where  $\vec{\wp}^{\,2} = \sum_{k=1}^{3}\wp_{k} \wp^{k}\,$.
\\
  \\
{\bf The determination of the square of the particle mass}: As the channel information capacity $I$ is from the definition the sum of the expected values (see (\ref{Fisher info dla theta n gradient}) and (\ref{pojemnosc C})),  the squared {\it mass} of a particle, defined as \cite{Frieden}:
\begin{eqnarray}
\label{m E p} 
m^{2} := \frac{1}{N\, c^{2}} \int_{\cal P}  d^{4} {\bf p} \sum_{n=1}^{N} |\tilde{q}_{n}({\bf p})|^{2} \,(\frac{E^{2}}{c^{2}}-\vec{\wp}^{\,2}\,) \;
\end{eqnarray}
is constant and does not depend on the statistical fluctuations of the energy $E$ and momentum  $\vec{\wp}$, i.e. at least as the mean after the integration is performed. 
Thus, for a free particle we can record (\ref{I by Fourier tr}) as follows: 
\begin{eqnarray}
\label{I by Fourier to m}
I\left[q({\bf x})\right] = I\left[\tilde{q}({\bf p})\right] = 4 N \, (\frac{m\, c}{\hbar})^{2} = const \; .
\end{eqnarray}
Therefore, we observe that the causality relation (\ref{przyczynowosc}) entails $I \geq 0$ in (\ref{informacja Stama vs pojemnosc informacyjna}) and then,  
in accord with (\ref{I by Fourier to m}) the condition $m^{2} \geq 0$, i.e. the lack of tachions in the theory \cite{dziekuje za neutron}. 
It must be pointed out that, according to (\ref{m E p}), the zeroing of particle mass would be impossible for the Euclidean space-time metric (\ref{metryka E}).  \\
\\
{\bf The Fourier information with $q$}: The condition  (\ref{I by Fourier to m}) means that: 
\begin{eqnarray}
\label{free field eq all 1}
K_{F} \equiv I\left[q({\bf x})\right] - I\left[\tilde{q}({\bf p})\right]=0 \; ,
\end{eqnarray}
what, using the constancy of  $4N(\frac{m\, c}{\hbar})^{2}$ and (\ref{norm condition}) can be rewritten as the condition fulfilled by the free field of the rank  $N$:
\begin{eqnarray}
\label{free field eq all 2}
K_{F} = \int_{\cal X} d^{4}{\bf x} \, k_{F} = 0 \; .
\end{eqnarray}
Here, the quantity $K_{F}$ defines the so-called {\it Fourier information} ($F$), where its density $k_{F}$  is equal to:
\begin{eqnarray}
\label{gestosc kF inf Fouriera}
k_{F} = 4 \,  \sum_{n=1}^{N} \left[\;\sum_{\nu=0}^{3}  \frac{\partial q_{n}({\bf x})}{\partial {\bf x}_{\nu}} \; \frac{\partial q_{n}({\bf x})}{\partial {\bf x}^{\nu}} - (\frac{m\, c}{\hbar})^{2} \, q_{n}^{2}({\bf x}) \right]  \; .
\end{eqnarray} 
{\bf Remark~5}. Ignoring the fact that from the above calculations $m^2$ emerges as the mean (\ref{m E p}), 
the equation (\ref{free field eq all 1}), and consequently  (\ref{free field eq all 2}), is the reflection of the Parseval's theorem 
and as such it is a tautology. 
This means that 
the Fourier transformation reflects the change of the basis in the statistical space  ${\cal S}$ only. Therefore, by itself only the condition (\ref{free field eq all 2}) does not superimpose any new 
constraint {\it unless $4 N \,(\frac{m\, c}{\hbar})^{2}$ is additionally determined as the structural information 
of the system, which in this particular case 
defines the type of the field, the scalar one} \cite{Frieden,Dziekuje informacja_1,Dziekuje informacja_2,Dziekuje za skrypt,Dziekuje za zasadyJJ}.

\section{Conclusions}

\label{conclusion}

A system without a structure dissolves itself, hence its
equation of motion requires besides the kinematical term (which is the channel information capacity $I$) a structural one. 
Intuitively, "during putting the structure upon" the information on the system has to be  maximized and this is performed by the addition of the structural information $Q$, which acts under the condition of zeroing of the {\it observed structural information principle} \cite{Dziekuje informacja_1,Dziekuje informacja_2,Dziekuje za skrypt,Dziekuje za zasadyJJ}.
Both the analysis of the analyticity of the log-likelihood function \cite{Dziekuje informacja_2} and the 
Rao-Fisher metricity of the statistical (sub)space ${\cal S}$  \cite{Amari Nagaoka book} allow for the formal construction of this observed differential form of the structural information principle \cite{Dziekuje informacja_2,Dziekuje za zasadyJJ}, which, when lifted to the expected level, connects the channel information capacity $I$ with the structural information $Q$.
However, when this structural differential constraint is established then from all distributions in ${\cal S}$ the one which minimizes total physical information $K \equiv I + Q$ is chosen. It is achieved via the {\it variational principle} $\, I + Q \rightarrow min \, $. \\ 
The above two information principles, i.e. firstly, the observed and expected structural information principle and secondly, the variational one, form the core of the EPI method of estimation \cite{Dziekuje informacja_2,Dziekuje za skrypt,Dziekuje za zasadyJJ} (see Introduction). In order to estimate correctly the distribution of the system, the EPI method uses these principles, estimating the equation of motions for the particular field theory model, or the equation generating the distribution in the particular model of the statistical physics. 
The whole procedure depends greatly on the form of the channel information capacity $I$ used in the particular physical model, which has to take into account the geometrical and physical preconditions, among these the metric of the base space ${\cal Y}$ and the (internal) symmetries of the physical system.  For example, different representations of the Lorentz transformation, which is the isometry of $I$, lead to its specific decomposition giving Klein-Gordon or Dirac equation \cite{Frieden,Dziekuje za skrypt,Dziekuje za models building}. Nonetheless, we cannot rule out the possibility that the opposite is also indisputable. For instance, from the basic property of the non-negativity of $I$, (\ref{informacja Stama vs pojemnosc informacyjna}), 
preceded by the causality  property (\ref{przyczynowosc}), which is
the statistical requirement of any successful observation, there  follows the very important physical property (\ref{I by Fourier to m}) of the non-existence of tachion  in the field theory models. 
\\
In accord with {\it Remark~5} underneath (\ref{gestosc kF inf Fouriera}), only when the structural information principle \cite{Dziekuje informacja_2,Dziekuje za skrypt,Dziekuje za zasadyJJ} fulfils also the mass relation (\ref{m E p}) which is put as the constrain upon the Fourier transformation (\ref{Fourier transf}), the entanglement\footnote{It follows from 
Section~\ref{Informacja Fouriera} that the Fourier transformation forms a kind of self-entanglement (between two representations of $I$, one of positions and the other of momentums) of the realized values of the variables of the system \cite{Frieden} and leads to (\ref{I by Fourier tr}). Nevertheless, without putting the structural information principle on the system, which is the analyticity requirement of the log-likelihood function and without the Rao-Fisher metricity of ${\cal S}$, the relation (\ref{I by Fourier tr}) remains a tautological sentence only. Indeed, in the case of the relation discussed in (\ref{I by Fourier tr}), the Fourier transformation has to relate the Fisher information matrix, which is in the second order term of the Taylor expansion of the log-likelihood function  \cite{Dziekuje informacja_2}  with remaining parts of this expansion.
} 
among the momentum degrees of freedom in the system appears. \\ 
Now, let us discuss a massless particle, e.g. photon. 
If, in agreement with Section~\ref{Maxwell field}, the condition (\ref{free field eq all 2})-(\ref{gestosc kF inf Fouriera}) is  adopted for a particle with a mass  $m=0\,$  and the amplitude is 
interpreted according to (\ref{amplitudy dla Maxwella}) and (\ref{A normalization}) as characterizing the photon, 
it would then necessitate answers to both  the question on the nature of the photon wave function \cite{Roychoudhuri} and the signature of the space-time metric from experiments.  Indeed, in the Minkowski space-time metric (\ref{metryka M}), according to (\ref{m E p}), the only possibility for a particle to be massless is that it fulfills the condition $E^{2}/c^{2}-\vec{\wp}^{\,2} = 0$ for 
all its Fourier monochromatic modes, if only the Fourier modes of massless particles are to possess a physical interpretation \cite{Dziekuje za skrypt}. 
However, this condition means that the Fourier modes are not entangled (in contrast to the massive particle case) and in principle,  the possibility to detect every individual mode should exist. If the individual Fourier mode frequencies of a light pulse were not detected,  it would have meant that they are not physical objects \cite{Roychoudhuri}. This would lead to a serious problem for the quantum interpretation of a photon as the physical realization of a particular Fourier mode. This problem recently  occurred in the light beam experiments discussed in \cite{Roychoudhuri}, which  suggests that the Fourier decomposed frequencies of a pulse do not represent actual optical frequencies, 
possibly implying that a real photon is a "lump of electromagnetic substance" without Fourier decomposition   \cite{Roychoudhuri,Dziekuje_Jacek_nova_2}. \\
Finally, in general, the channel information capacity and the structural information form the basic concepts in the analysis of  entanglement in the system. 
Nonetheless, e.g. for analyses of this phenomenon for the fermionic particle, the complex amplitude should sometimes be introduced \cite{Frieden,Dziekuje za skrypt}.

\section*{Acknowledgments}
The contribution of J. Syska to the work has been supported by L.~J.~Ch. \\ 
The work of Jan S{\l}adkowski has been supported by the project {\bf Quantum games: theory and implementations} financed by  
the {\bf National Science Center} under the contract no {\bf DEC-2011/01/B/ST6/07197}, which has supported also J.~Syska in part.

\end{document}